\documentclass[aps,nofootinbib,showpacs,twocolumn]{revtex4}

\begin{document}

\preprint{gr-qc/0403103}

\title{Quasilocal Center-of-Mass}

\author{James M. Nester}\email{nester@phy.ncu.edu.tw}
\affiliation{Department of Physics and Institute of Astronomy,
National Central University, Chungli 320, Taiwan}

\author{Feng-Feng Meng}
\author{Chiang-Mei Chen}\email{cmchen@phys.ntu.edu.tw}

\affiliation{Department of Physics, National Central University,
Chungli 320, Taiwan}

\date{\today}


\begin{abstract}
Gravitating systems have no well-defined local energy-momentum
density. Various quasilocal proposals have been made, however the
center-of-mass moment (COM) has generally been overlooked.
Asymptotically flat graviating systems have 10 total conserved
quantities associated with the Poincar{\'e} symmetry at infinity.
In addition to energy-momentum and angular momentum (associated
with translations and rotations) there is the boost quantity: the
COM. A complete quasilocal formulation should include this
quantity. Getting good values for the COM is a fairly strict
requirement, imposing the most restrictive fall off conditions on
the variables. We take a covariant Hamiltonian approach,
associating Hamiltonian boundary terms with quasilocal quantities
and boundary conditions. Unlike several others, our {\it covariant
symplectic} quasilocal expressions do have the proper asymptotic
form for all 10 quantities.
\end{abstract}

\pacs{04.20.-q,04.20.Fy}

\maketitle

\section{Introduction}

Associated with the flat spacetime geometric symmetries there are
10 conserved quantities: energy-momentum (EM) (translations),
angular momentum (AM) (rotations) and its often overlooked
covariant partner, the center-of-mass moment (COM) (boosts).
Asymptotically flat gravitating systems have all of these
quantities.  For such spacetimes the total values exist globally
but there are no well defined local densities.

The localization of energy-momentum for gravitating systems
remains an outstanding problem. The source of gravity is the EM
density. EM is exchanged {\it locally} between sources and
gravity, hence we expect something like a local gravitational EM
density.  But standard techniques give only non-covariant
(coordinate dependent) {\it pseudotensors}; gravity has no proper
EM (nor AM/COM) density. This is consistent with the equivalence
principle: gravity cannot be detected at a point. It is now
believed that the proper idea is {\it quasilocal} quantities
(i.e., associated with a closed 2-surface). We want good
quasilocal expressions for EM and AM/COM \cite{Ne04}.

Most earlier quasilocal investigations focused on energy-momentum.
Angular momentum also received some attention, however its
4-covariant associate, the center-of-mass moment, has been very
much neglected. Obtaining the correct value for this quantity is
actually quite a severe requirement; consequently it provides a
very discriminating test for proposed expressions.

A good approach to energy-momentum is via the Hamiltonian. The
Hamiltonian for a (finite or infinite) region necessarily includes
a boundary term.  The  Hamiltonian boundary terms give both the
quasilocal quantities and the  boundary conditions. We have
developed a covariant Hamiltonian formalism which has given
certain special {\it covariant-symplectic} boundary terms. These
expressions have already been well tested for EM and AM
\cite{Chen:1998aw,Chen:1994qg,Chang:1999da,Chen:2000xw}.

The COM test has recently been applied to our
``covariant-symplectic'' Hamiltonian-boundary-term quasilocal
expressions \cite{Meng}. Here we briefly describe the results of
that investigation,  their correspondence with the asymptotically
acceptable total expressions, outline  how they give the
quasilocal COM for Einstein's GR, and compare them with other
quasilocal expressions.

\section{The Hamiltonian Approach}
The Hamiltonian depends on a spacetime displacement vector field;
it includes both a volume term and a spatial boundary term:
\begin{equation}
H(N)=\int_{\Sigma}{\cal H}(N) =\int_{\Sigma}N^\mu{\cal
H}_\mu+\oint_{S=\partial\Sigma}{\cal B}(N).
\end{equation}
The Hamiltonian boundary term has important roles. The value of
the Hamiltonian gives the quasilocal quantities. This value is
conserved; different displacements give the quasilocal EM and
AM/COM.  Since ${\cal H}_\mu$ vanishes ``on shell'' the value of
the Hamiltonian is determined by the {\it Hamiltonian boundary
term} (HBT): ${\cal B}(N)$. The ``natural'' HBT inherited from the
Lagrangian can---{\it and should}---be adjusted (as first clearly
noted for GR by Regge and Teitelboim \cite{RT}).

The general expression for ${\cal B}$ depends on the choice of
variables, a displacement vector field (e.g. translation for
energy-momentum and rotation for angular momentum), a reference
configuration and boundary conditions. Boundary conditions are
determined by the {\it Hamiltonian variation boundary principle}.
Formally there are an infinite number of possible choices for
${\cal B}$ corresponding to different reference configurations and
boundary conditions, hence selection criteria are needed. Good
asymptotics is one important condition. Another is {\it
covariance}.  We found that there are only two choices which give
rise to a boundary term in the variation of the Hamiltonian which
requires us to hold fixed (certain projected components of) {\it
covariant} objects (essentially these two choices correspond to
Dirichlet or Neumann boundary conditions). In particular for GR,
associated with boundary conditions imposed on
$\pi^{\mu\nu}:=(-g)^{1/2}g^{\mu\nu}$ we have
\begin{equation}
{\cal B}_g^{\mu\nu}(N) = N^{\tau}\pi^{\beta\lambda} \Delta
\Gamma^\alpha{}_{\beta\gamma}
\delta^{\mu\nu\gamma}_{\alpha\lambda\tau} + {\bar D}_\beta
N^\alpha \Delta \pi^{\beta\lambda}
\delta^{\mu\nu}_{\alpha\lambda}.
\end{equation}
Here $\Delta \Gamma:=\Gamma-\bar\Gamma$, $\Delta \pi:=\pi-\bar
\pi$, with $\bar\Gamma$, $\bar \pi$ being reference values. (The
reference values have a simple meaning: all quasilocal quantities
vanish when the dynamic variable takes on the reference values;
the standard reference choice is flat Minkowski spacetime).
Technically we prefer the differential form version:
\begin{equation}
{\cal B}(N) = \Delta \Gamma^{\alpha\beta} \wedge i_N
\eta_{\alpha\beta} + {\bar D}^{[\beta} N^{\alpha]} \Delta
\eta_{\alpha\beta},\label{B}
\end{equation}
where
$\eta^{\alpha\beta}:=*(\vartheta^\alpha\wedge\vartheta^\beta)$.
For details see
\cite{Chen:1998aw,Chen:1994qg,Chang:1998wj,Chang:1999da,Chen:2000xw}.

Note the form of these Hamiltonian boundary terms:
\begin{equation}
B(N) = N^\mu {\cal P}_\mu + D^{[\mu} N^{\nu]} {\cal
S}_{\mu\nu},\label{Bgr}
\end{equation}
qualitatively: ${\cal B}$=``Freud'' + ``Komar''; for AM it is easy
to see that this corresponds to ``orbital'' +``spin''. Let the
displacement have the asymptotic Poincar{\'e} form $N^\mu = Z^\mu
+ \omega^\mu{}_\nu x^\nu$, with $\omega^{\mu\nu} =
\omega^{[\mu\nu]}$. Then
\begin{equation}
{\cal B} = Z^\mu {\cal P}_\mu + {1\over2} \omega_{\mu\nu} {\cal
J}^{\mu\nu},\label{Bpoin}
\end{equation}
where
\begin{equation}
{\cal J}^{\mu\nu} = x^\mu {\cal P}^\nu - x^\nu {\cal P}^\mu +
{\cal S}^{\mu\nu}.\label{J}
\end{equation}
This includes angular momentum: $x^i {\cal P}^j - x^j {\cal P}^i$,
and the center-of-mass: $x^0 {\cal P}^k - x^k {\cal P}^0$.

\section{Results}
We have compared our expressions with various other proposals. We
first consider total expressions at spatial infinity. MTW
\cite{MTW}, Eq.~(20.9), gives the necessary asymptotic form for
all 10 Poincar{\'e} quasilocal quantities (in an essentially
4-covariant form). It is straightforward to verify  that our
expressions (\ref{B}--\ref{J}) have that asymptotic limit; the
$DN$ contribution, ${\cal S}$, is distinctive \cite{Meng}. We
have, in the first part of Table I, summarized the degree of
success for various pseudotensor expressions, including those of
Einstein/Freud, Duan and Feng, Landau and Lifshitz, Weinberg/MTW,
Papapetrou and Goldberg
\cite{Chang:1998wj,Chang:1999da,Chen:2000xw,Vu}. Turning to
Hamiltonian approaches, we note that the famous ADM expressions
\cite{ADM} have no $DN$ term, these investigators did not consider
the COM. This was done later by Regge and Teitelboim (RT)
\cite{RT}; their Eq.~(5.13) has a $DN$ term which plays an
important role in determining the COM. Beig and \'O Murchadha
(B{\'o}M) \cite{BoM} have given a refinement of the RT work (see
their Eq. (3.37)); they noted that an explicit reference
configuration is needed. More recently Szabados
\cite{Szabados:2003yn} has given an even more careful discussion,
further refining the B{\'o}M results. These investigations have
shown the overall importance of the COM: in order for it to be
well defined one must impose the most strict asymptotic conditions
of the variables.  The second part of Table I. provides a summary.

\begin{table}
\begin{center}
\caption{Success of asymptotic total expressions}
\begin{tabular}{p{5cm}ccc}
 & EM & AM & COM \\
\hline
 {\bf Pseudotensor approaches:} & & & \\
 Einstein/Freud & ok & no & no \\
 Duan \& Feng & ok & \multicolumn{2}{c}{special gauge} \\
 LL & ok & ok & ok \\
 Weinberg/MTW & ok & ok & ok \\
 Papapetrou & ok & ok & ok \\
 Goldberg & ok & ok & ok \\
\hline
 {\bf Hamiltonian formulations:} & & & \\
 ADM & ok & ok & no \\
 RT & ok & ok & good \\
 B{\'o}M & ok & ok & better \\
 Szabados & ok & ok & best \\
 \hline
\end{tabular}
\end{center}
\end{table}

Turning to quasilocal proposals, the seminal Brown and York work
\cite{Brown:1992br} has no $DN$ term---and no COM discussion (a
serious shortcoming in our view).  Both the Witten spinor
Hamiltonian and Tung's spin 3/2 Hamiltonian have no $DN$ term, and
apparently cannot give AM/COM \cite{ca6p}. The apparently
necessary $DN$ terms do appear in our covariant symplectic
expressions (\ref{B}). (Note: one of our ``covariant symplectic''
expressions was found independently by Katz, Bi{\v c}{\'a}c and
Lynden-Bell \cite{Katz:1996nr}).

This brings us to our central question: is the $DN$ term
absolutely essential? We find that it is necessary not only for
the Hamiltonian variation but also generally for the COM value.
Indeed the COM value is the only case for which it plays an
essential role. The reason can be understood by first noting that
asymptotically, because of the fall off rates, this term can
affect only AM and/or COM, not EM.

Consider first angular momentum.  We found that the $DN$
contribution can play an important role for angular momentum, but
it is not essential.  In particular Vu considered the teleparallel
equivalent of GR, GR${}_{||}$. Its tetrad version GRtet, which
lacks $DN$ terms, can give the AM---but only in a certain frame
gauge, whereas the GR${}_{||}$ version (which has $DN$ terms)
succeeds in a general frame \cite{Vu}. On the other hand, many
investigations have obtained the correct angular momentum at
spatial infinity without using such a term
\cite{RT,BoM,Szabados:2003yn}; also one can likewise get good
quasilocal angular momentum  without such a contribution
\cite{Brown:1992br}.

We find, however, that $DN$ terms  play an essential role in
obtaining the COM.  It is readily apparent from the investigations
at spatial infinity \cite{RT,BoM,Szabados:2003yn} that $DN$
contributes only to the COM. Ho found that the $DN$ terms are
essential to get the COM in the GR${}_{||}$ theory
 \cite{Ho}.
Here, for GR, we outline a calculation showing their important
role in obtaining the correct COM.

 We did a simple test on the
eccentric Schwarzschild geometry. Take the isotropic Schwarzschild
solution,
\begin{equation}
ds^2 = -N^2 dt^2 + \varphi^4 (dx^2 + dy^2 + dz^2),
\end{equation}
where $\varphi=1+m/2r$ and $N\varphi=1-m/2r$, and displace the
center:
\begin{equation}
{1\over r} \to {1\over |{\bf r}-{\bf a}|} \simeq {1\over r}+ {{\bf
a\cdot r}\over r^3}.
\end{equation}
Now, using the obvious Minkowski reference, evaluate (\ref{B})
with $N^0={\bf v\cdot r}$, this is a ``boost'' in the $\bf v$
direction. The ``Freud'' term gives: $ (2/3)m{\bf a\cdot v}$ and
the ``Komar'' term gives $(1/3)m{\bf a\cdot v}$. Together they
give the total center-of-mass moment: $ m{\bf a}$.

\begin{table}
\begin{center}
\caption{Success of various quasilocal expressions}
\begin{tabular}{p{5cm}ccc} Quasilocal expressions:
 & EM & AM & COM \\
\hline
%
 Witten spinor Hamiltonian & ok & no & no \\
 Tung's 3/2 QSL Hamiltonian & ok & no & no \\
 Brown \& York & ok & ok & ok \\
 GRtet & ok & special & incorrect \\
  & &  gauge & \\
 GR${}_{||}$ & ok & ok & ok \\
 covariant-symplectic GR & ok &
ok & ok \\
 \hline
\end{tabular}
\end{center}
\end{table}

The popular Brown and York (BY) quasilocal formalism
\cite{Brown:1992br} appears to have a major shortcoming: according
to our discussion, from their quasilocal energy expression:
\begin{equation}N(k-k_0),\label{BYe}\end{equation}
 it seems like one cannot get the correct total COM, since there is no $DN$ term.
However Baskaran, Lau and Petrov \cite{Baskaran:2003pk} have
demonstrated, via a remarkable elaborate calculation, that
(\ref{BYe}) asymptotically {\it agrees} (up to a term with
vanishing integral) with the B{\'o}M expression.  Thus the BY
expression, contrary to our original belief, does give the correct
COM value. In Table II we summarize the success of the various
quasilocal expressions (for a detailed discussion of the Witten
spinor Hamiltonian and Tung's spin 3/2 QSL Hamiltonian see
\cite{ca6p}; for GRtet and GR${}_{||}$, see \cite{Vu,Ho,mgXcmtp}).

\bigskip

\section{Conclusion}

Asymptotically flat spaces have 10 conserved quantities. A good
description of EM should also include AM/COM (note: relativistic
invariance requires COM along with AM). Many proposals have,
unfortunately, overlooked the COM. To get the correct COM is a
strong requirement; the COM imposes the strictest fall off
conditions. Considering the COM can be decisive. From COM
consideration we find that tetrad GR, Witten spinor, and Tung's
spin 3/2 quasilocal proposals all have {\it serious shortcomings}.
Aside from the BY expression, only our covariant-symplectic GR,
and GR${}_{||}$ satisfy the good COM requirement. Their asymptotic
spatial limit does give the correct value for the total COM.
Moreover the asymptotic form of our covariant symplectic
Hamiltonian boundary expressions agrees with accepted expressions
\cite{MTW,RT,BoM,Szabados:2003yn}.  To our knowledge they are the
best behaved expressions which have so far been identified.

In summary, our investigation considered an important neglected
quantity: the quasilocal COM; the investigation provides
additional support for the covariant symplectic quasilocal
expression.


\section*{Acknowledgments}

We especially wish to thank S. R. Lau and A. N. Petrov for their
help in correcting our misunderstandings of their remarkable work
\cite{Baskaran:2003pk} and the Brown-York quasilocal energy. This
work was supported by the National Science Council of the ROC
under grant number NSC92-2112-M-008-050, NSC92-2119-M-008-024.


\begin{references}

\bibitem{Ne04} J.~M. Nester,
Class.\ Quantum Grav. {\bf 21}, S261 (2004).

\bibitem{Chang:1999da}
C.~C.~Chang, J.~M.~Nester and C.~M.~Chen,
 ``Energy-Momentum (Quasi-)Localization for Gravitating Systems,''
in {\sl Gravitation and Astrophysics}, ed L. Liu et al (World
Scientific, Singapore, 2000) 163, [arXiv:gr-qc/9912058].


\bibitem{Chen:1994qg}
C.~M.~Chen, J.~M.~Nester and R.~S.~Tung,
Phys.\ Lett.\ A {\bf 203}, 5 (1995) [arXiv:gr-qc/9411048].

\bibitem{Chen:1998aw}
C.~M.~Chen and J.~M.~Nester,
Class.\ Quant.\ Grav.\  {\bf 16}, 1279 (1999)
[arXiv:gr-qc/9809020].

\bibitem{Chen:2000xw}
C.~M.~Chen and J.~M.~Nester,
Grav.\ Cosmol.\  {\bf 6}, 257 (2000) [arXiv:gr-qc/0001088].

\bibitem{Meng}
F.~F.~Meng, ``Quasilocal Center of Mass Moment for GR (MSc.
thesis, NCU, 2002)

\bibitem{RT} T.~Regge and C.~Teitelboim,
Annals Phys.\ {\bf 88}, 286 (1974).

\bibitem{Chang:1998wj}
C.~C.~Chang, J.~M.~Nester and C.~M.~Chen,
Phys.\ Rev.\ Lett.\  {\bf 83}, 1897 (1999) [arXiv:gr-qc/9809040].

\bibitem{MTW}
  C.~W.~Misner, K.~S.~Thorne and J.~A.~Wheeler,
  {\sl Gravitation},
  (Freeman, San Francisco, 1973).

\bibitem{Vu} K.~H.~Vu, ``Quasilocal Energy-Momentum and Angular
Momentum for Teleparallel Gravity'' (MSc.~Thesis, NCU, 2000)

\bibitem{ADM} A.~Arnowit, S.~Deser and C.~W.~Misner, The dynamics
of general relativity, in {\it Gravitation: An Introduction to
Current Research}, ed L.~Witten (Wiley, New York, 1962) pp
227--65.


\bibitem{BoM}
R. Beig  and N. \'O Murchadha,
%
Annals Phys.\ {\bf 174}, 463 (1987).

\bibitem{Szabados:2003yn}
L.~B.~Szabados,
Class.\ Quant.\ Grav.\  {\bf 20}, 2627 (2003)
[arXiv:gr-qc/0302033].


\bibitem{Brown:1992br}
J.~D.~Brown and J.~W.~.~York,
Phys.\ Rev.\ D {\bf 47}, 1407 (1993) [arXiv:gr-qc/9209012].




\bibitem{ca6p} C.~M. Chen, J.~M.~Nester and R.~S.~Tung,
``Spinor Formulations for Gravitational Energy-Momentum'', in {\it
Clifford Algebras: Applications to Mathematics, Physics and
Engineering (Progress in Mathematical Physics} vol 34), ed
R.~Ablamowicz (Birkhauser, Boston, 20003) pp 417--30
[arXiv:gr-qc/0209100].

\bibitem{Katz:1996nr}
J.~Katz, J.~Bi{\v c}{\'a}k and D.~Lynden-Bell,
Phys.\ Rev.\ D {\bf 55}, 5957 (1997).

\bibitem{Ho}
 F.~H.~Ho, ``Quasilocal center-of-mass for
GR${}_{||}$'' (MSc.~thesis, NCU, 2003).


\bibitem{Baskaran:2003pk}
D.~Baskaran, S.~R.~Lau and A.~N.~Petrov,
Annals Phys.\ {\bf 307}, 90 (2003) [arXiv:gr-qc/0301069].

\bibitem{mgXcmtp} J.~M.~Nester, F.~H.~Ho and C.~M.~Chen,
``Quasilocal Center-of-Mass for Teleparallel Gravity'', to appear
in {\it Proceedings of the 10th Marcel Grossman meeting} (Rio de
Janeiro, 2003) [arXiv:gr-qc/0403101].














\end{references}
\end{document}